\documentclass[journal=langmuir,manuscript=article]{achemso}

\usepackage{color}

\usepackage[version=3]{mhchem} 

\usepackage{graphicx}
\usepackage{subcaption}
\usepackage{relsize} 

\author{Desislava Dimova}
\affiliation[University of Sofia]
{Department of Atomic Physics, University of Sofia}
\author{Stoyan Pisov}
\affiliation[University of Sofia]
{Department of Atomic Physics, University of Sofia}
\email{pisov@phys.uni-sofia.bg}
\phone{+359 (0)2 8161828}
\author{Nikolay Panchev}
\affiliation[Bulgarian Academy of Sciences]
{Institute of Physical Chemistry}
\email{patcho75@yahoo.com}
\author{Miroslava Nedyalkova}
\affiliation[University of Sofia]
{Department of General and Inorganic Chemistry, University of Sofia}
\author{Sergio Madurga}
\email{s.madurga@ub.edu}
\affiliation[University of Barcelona]
{Material Science and Physical Chemistry Department and IQTCUB}
\author{Ana Proykova}
\email{anap@phys.uni-sofia.bg}
\affiliation[University of Sofia]
{Department of Atomic Physics, University of Sofia}

\title[Article title]
  {A model provides insight into electric field-induced rupture mechanism of water-in-toluene emulsion films
}

\abbreviations{IR,NMR,UV}

\begin{document}



\begin{abstract}
This paper presents the first MD simulations of a model, which we have designed for understanding the development of electro-induced instability of a thin toluene emulsion film in contact with saline aqueous phase. This study demonstrates the charge accumulation role in toluene film rupture when a DC electric field is applied. The critical value of the external field at which film ruptures, thin film charge distribution, capacitance, number densities and film structure have been obtained in simulating the system within $NVT$ and $NPT$ ensembles.  A mechanism of thin film rupture driven by the electric discharge is suggested.We show that $NPT$ ensemble with a constant surface tension is a better choice for further modeling of the systems that resemble more close the real films.
\end{abstract}

\section{Introduction}
Water-in-oil emulsions are commonly formed during petroleum production and pose serious threats to installations and quality of the final product. The electrical phase separation has been used in the petroleum industry for separating water-in-crude oil dispersion's by applying a high electric field onto the flowing emulsion to affect flocculate and coalescence of dispersed water droplets \cite{Cottrell1991,eow02}.
It has been realized that the emulsion is stabilized by a thin film formed between two drops when approaching each other. Thus demulsification requires rupturing of this thin liquid film. 
Generally, the main purpose of an applied electrical field is to promote contact between the drops and to help in drop\textendash drop coalescence. Pulsed DC (direct current) and AC (alternative current) electric fields are preferred over constant DC fields for efficient coalescence. Recent studies have helped to clarify important aspects of the process such as partial coalescence and drop\textendash drop non-coalescence but key phenomena such as thin film breakup and chain formation are still unclear \cite{Mhatre2015}.
Despite of the tremendous practical importance of enhanced coalescence, the mechanism of separation  is not fully understood \cite{Isaacs01} beyond the  perception that the electrical force facilitates the coalescence between small drops.\\
To help in understanding the inherent processes, computational models were designed  to simulate coalescense of droplets under realistic experimental conditions. Molecular dynamics (MD) method is an useful tool for the purpose.  Koplik and Banavar \cite{Koplik2} did a pioneer work in modeling the coalescence of two Lennard-Jones liquid droplets in a second immiscible fluid using MD simulations. The authors found that coalescence of liquid droplets  was completely driven by van der Waals and electrostatic interactions when the velocities of the droplets were small. The coalescence began when the molecules on the boundary of one droplet thermally drifted to the range of attraction of the other droplet and formed a string to attract both sides of the molecules. \\
Zhao et al.\cite{Zhao2004} reported a MD study of the coalescence of two nanometer-sized water droplets in n-heptane, a system that is commonly encountered in the oil sands industry. Similarly, the coalescence process was initiated by the molecules at the edge of the clusters, which interacted with each other and formed a bridge between two clusters. Eventually, these molecules attracted and pulled out other molecules from their own respective cluster to interact with those from the other cluster. Authors made an important conclusion that the coalescence in n-heptane would occurred only if the two droplets were very close to each other ($\sim 0.5\, nm$). If they were far apart (e.g., $1\, nm$), external driving forces should be applied. 

However, experimental results for electrical properties and electric field-induced rupture of single thin films are scarce, which limits the comparison with computations to several measurable quantities - pore formation, and the critical voltage for film rupture.\\ Anklam et al.  \cite{Anklam1999} experimentally demonstrated that the electric-filed induced pore formation was the reason for break-up of emulsion films. Panchev et al. \cite{Panchev200874} 
developed a method allowing simultaneous investigation of a single water-in-oil emulsion film by both microinterferometry and electrical measurements. This method allows in a single experiment to measure the critical voltage of film rupture, the film thickness, the drainage rate, and the disjoining pressure laying the groundwork for computational studies. \\
 
In this paper we present computational results for pore formation and film rupture obtained with a model, which we have designed to imitate the rupture of the film under a step-wise increase of the electric field as it has been applied  in the experiment \cite{Panchev200874}. The film is immersed in a sodium chloride solution. In the model and also in the experiment, the electric field is applied perpendicularly to the film, which separates two water droplets.\\

The model of the thin film developed for the present study can be considered as a useful starting basis  for a further study of the stability and the structure of thick emulsion films that are stabilized by indigenous crude oil surfactants, namely asphaltenes, resins and naphthenic acids. It is worth mentioning that so far there is almost complete lack of understanding of the intimate structural details of the crude petroleum-like films. Therefore, current industrial practice of utilization of chemical additives in combination with electric field applications has for long time been widely viewed as a ``work of art''.

\section{The model and simulation procedure}

To accurately simulate the interfacial phenomena, we have applied the classical MD method for the case of two canonical ensembles - $NVT$  and  $NPT$. The choice of ensembles in MD simulations of finite-size systems has already been shown to play an important role in coexisting phases \cite{Pisov_2001}. MD Simulations provide detailed information on the molecular structure of the interface when the intermolecular potential is available \cite{Koplik,Koplik2014,Pisov_2012}.\\

The model system is a $5\, nm$ thick toluene film located perpendicularly to the $z$-axis  of the simulation box. The size of the box  $24.8 \times 24.8 \times 24.8\, nm$ ensures that no artifacts will appear when $3D$ periodic boundary conditions are implemented to diminish finite-size effects. The box contains also  water molecules and $Na^+$ and $Cl^-$ ions at a concentration of $1M$.  The force field parameters of the  ions $Na^+$ and $Cl^-$ included in the model are  taken from {\it Gromos96}  \cite{doi:10.1021/jp984217f}. Parameters for toluene molecules are derived from  benzyl side chain of phenylalanine molecule. Three-site $SPC$ (simple point charge) water model is used \cite{doi:10.1021/j100308a038}.

Large-scale molecular dynamics simulations of the model system are performed with the help of the  {\it GROMACS} package, designed to simulate the Newtonian equations of motion for systems with hundreds to millions of particles \cite{Berendsen199543}. The simulations are performed in  canonical $NVT$  and  $NPT$ ensembles which keep the total number of atoms constant; the temperature is $T = 298 K$.  In the $NVT$ ensemble the constant volume equals to the size of the simulation box, $24.8 \times 24.8 \times 24.8\, nm$. In the case of the $NPT$ ensemble the system is equilibrated at the constant pressure of $1\, bar$. After the equilibration, the simulation is performed at a constant surface tension $\gamma = 36.4\, mN/m$ between toluene and water \cite{Drelich_2002}. 
 In preliminary MD runs the  simulation time of  $5\, ns$ was determined to be sufficient for  thermodynamic equilibration of the total energy,  pressure, and temperature of the model system.\\
 An external electric field is applied  in the $z$ direction of the simulation box. In the $NVT$ ensemble the electric field strength is changed from $0$ to $120\, mV/nm$ in steps of $20\, mV/nm$,  while in the $NPT$ ensemble the strength is changed   from $0$ to $75\, mV/nm$ in steps of $25\, mV/nm$ .

\section{Results and discussion}
\subsection{$NVT$ simulation - build-up of interfacial charge}
\begin{figure}[ht]
\begin{center}
\includegraphics[width=.8\textwidth]{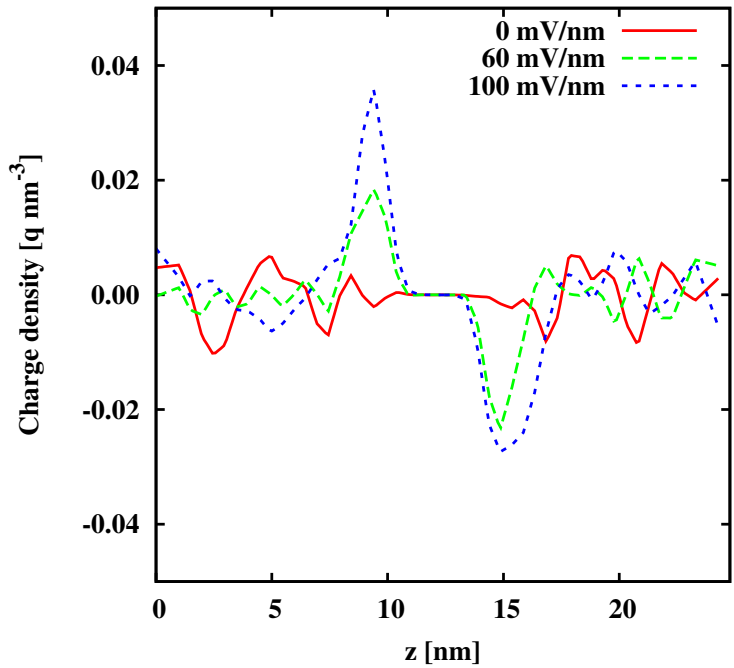}
\end{center}
\caption{Calculated free ion charge distribution in the $z$-direction demonstrates the build up of interfacial charge with the applied field increase - $0$, $60$ and $100$ $mV/nm$  }
\label{fig:nvt_charges_all_voltage}
\end{figure}

The interaction between the toluene film and the surrounding water molecules and ions results in a dynamic charge distribution. The Figure \ref{fig:nvt_charges_all_voltage} illustrates the ion (sodium and chlorine) charge density  distribution in the $z$-direction. The distribution is computed within the last $2.5\, ns$ of the run (total run time $5\, ns$).  The three curves correspond to  $0$, $60$ and $100\, mV/nm$ strengths of the external electric field. At $0\, mV/nm$ (red curve), which calibrates the results, the charge  fluctuates around zero at the film interfaces. A non-zero external field induces charge accumulation at film interfaces. The accumulated charge is drawn as peaks in the charge distribution. The curves for  $60\, mV/nm$  (green) and $100\, mV/nm$ (blue) electric field strengths show that the accumulated charge increases with the  field increase: one interface of the toluene film is charged positively due to $Na^+$ accumulation, while the other interface  is charged negatively due to  $Cl^-$ ion accumulation.  Thus, the emulsion film, subjugated to the external electric field, resembles charging of a parallel-plate capacitor. At all applied fields the ion charge fluctuates around zero away from the film.

\begin{figure}[h]
\centering
\begin{subfigure}{.45\textwidth}
  \centering
  \includegraphics[width=\textwidth]{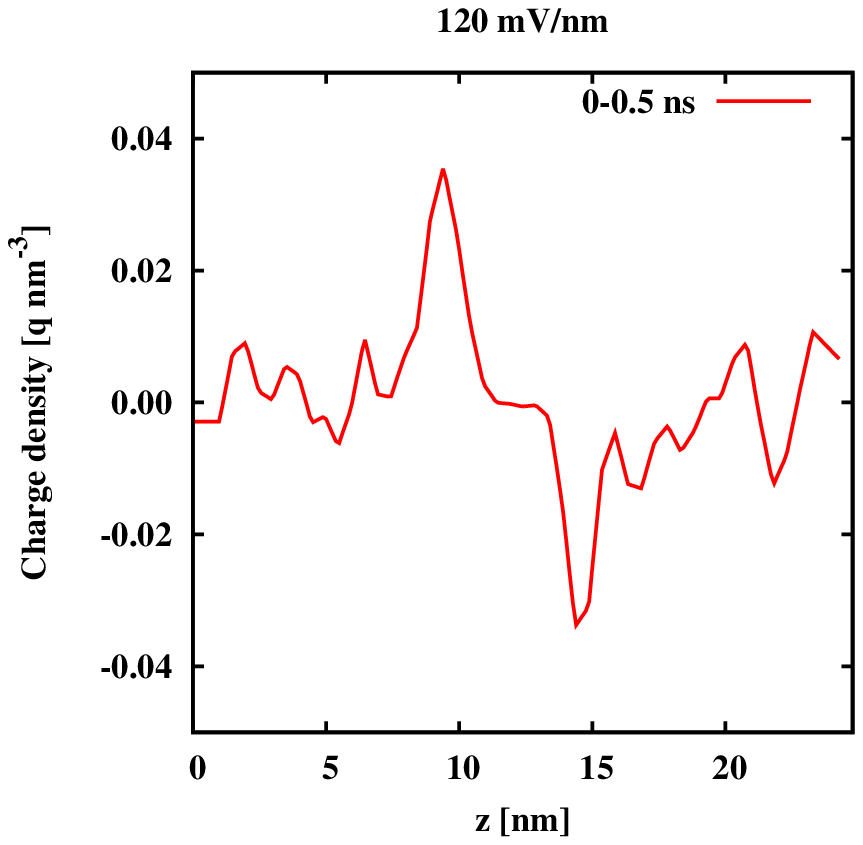}
  \caption{Free ions charge distribution at $120\, mV/nm$ between $0 - 0.5\, ns$}
  \label{fig:120mV_0-05ns}
\end{subfigure}%
\begin{subfigure}{.45\textwidth}
  \centering
  \includegraphics[width=\textwidth]{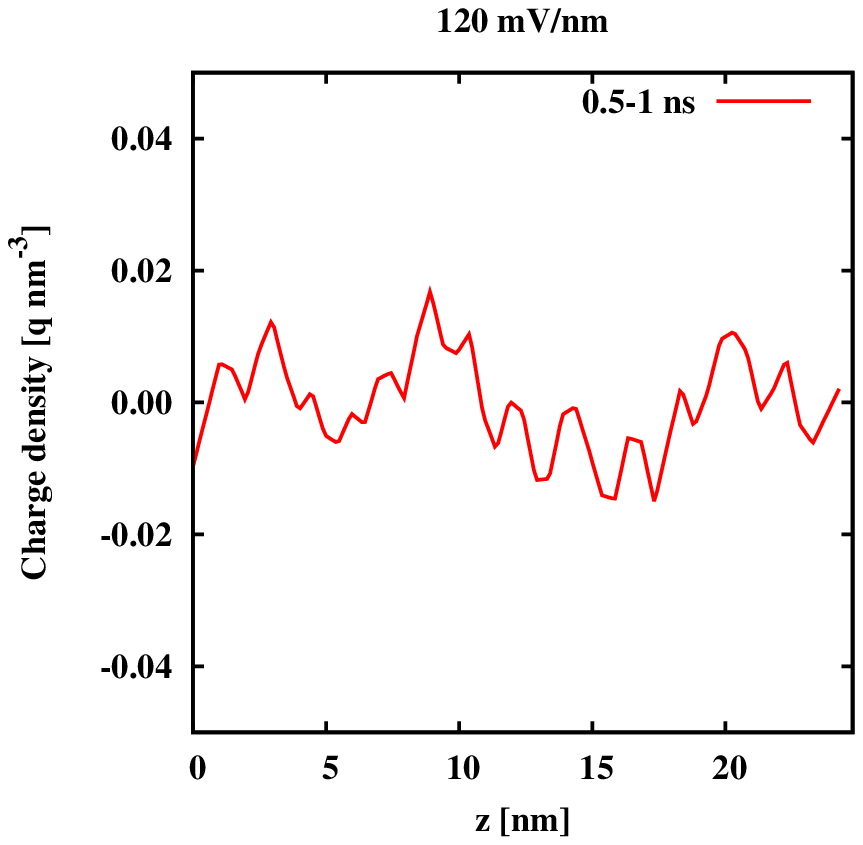}
  \caption{Free ions charge distribution at $120\, mV/nm$ between $0.5 - 1\, ns$}
  \label{fig:120mV_05-1ns}
\end{subfigure}
\begin{subfigure}{.45\textwidth}
  \centering
  \includegraphics[width=\textwidth]{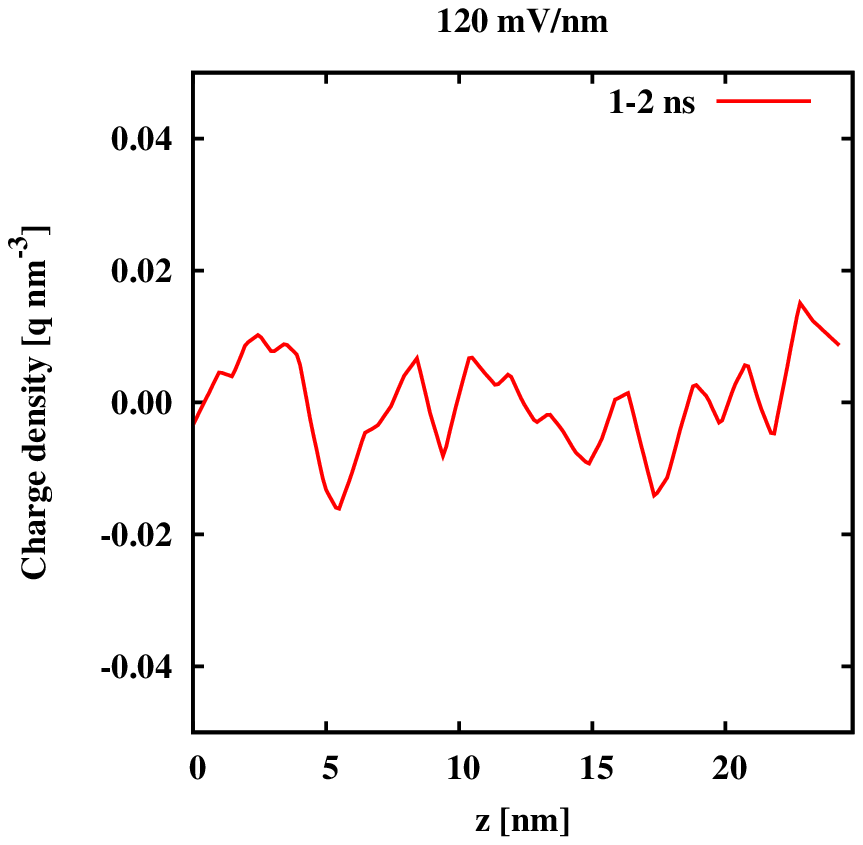}
  \caption{Free ions charge distribution at $120\, mV/nm$ between $1 - 2\, ns$}
  \label{fig:diff_120mV_1-2ns}
\end{subfigure}
\begin{subfigure}{.45\textwidth}
  \centering
  \includegraphics[width=\textwidth]{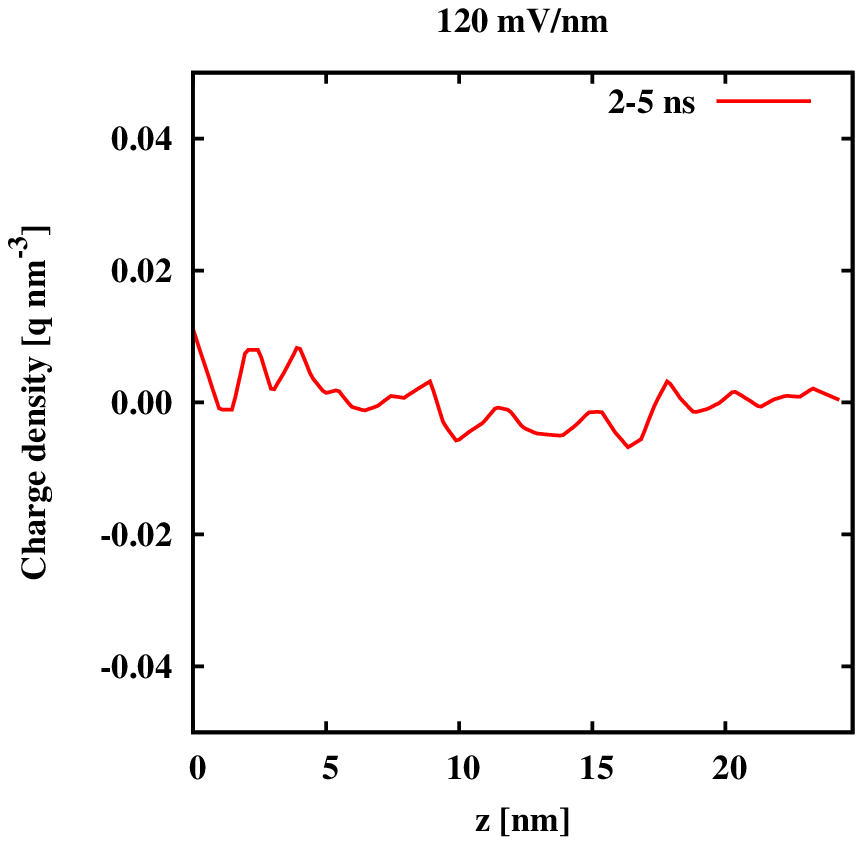}
  \caption{Free ions charge distribution at $120\, mV/nm$ between $2 - 5\, ns$}
  \label{fig:diff_120mV_2-5ns}
\end{subfigure}
\caption{Free ions charge distribution at $120\, mV/nm$}
\label{fig:nvt_charge_distrib_120mV}
\end{figure}

Averaging over time was performed over the last 2.5 $ns$ of the production run.
An external electric field with strengths up to $100\, mV/nm$ does not rupture the toluene film for  the duration of the $NVT$ simulation -  $5\, ns$. 
It should be noted that the  charge distribution $NVT$ obtained at the $120\, mV/nm$ field within time intervals less than $500\, ps$  shown in the Figure \ref{fig:120mV_0-05ns} feature on average the same patterns as the distribution computed   $100\, mV/nm$ field shown with a blue line in the Figure \ref{fig:nvt_charges_all_voltage}.\\
Electric discharge is initiated after the pore formation as it is seen in the Figure \ref{fig:diff_120mV_2-5ns}. After $2\, ns$, the accumulated interfacial charge is drained - no peaks in the charge distribution of the ruptured film. 

\begin{figure}[h]
\centering
\begin{subfigure}{.45\textwidth}
  \centering
  \includegraphics[width=.8\textwidth]{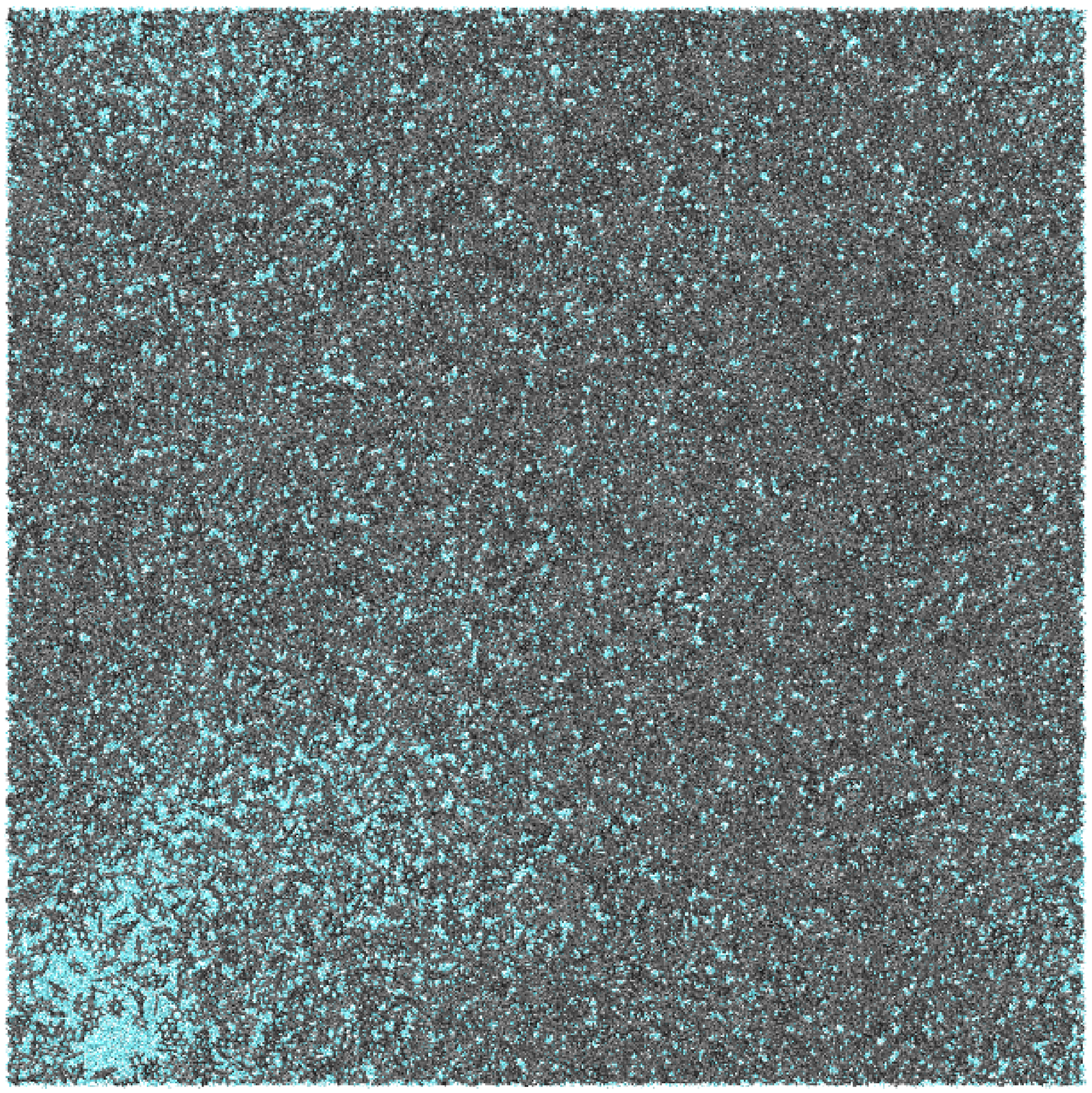}
  \caption{$500\, ps$}
  \label{fig:top_view_500ps}
\end{subfigure}%
\begin{subfigure}{.45\textwidth}
  \centering
  \includegraphics[width=.8\textwidth]{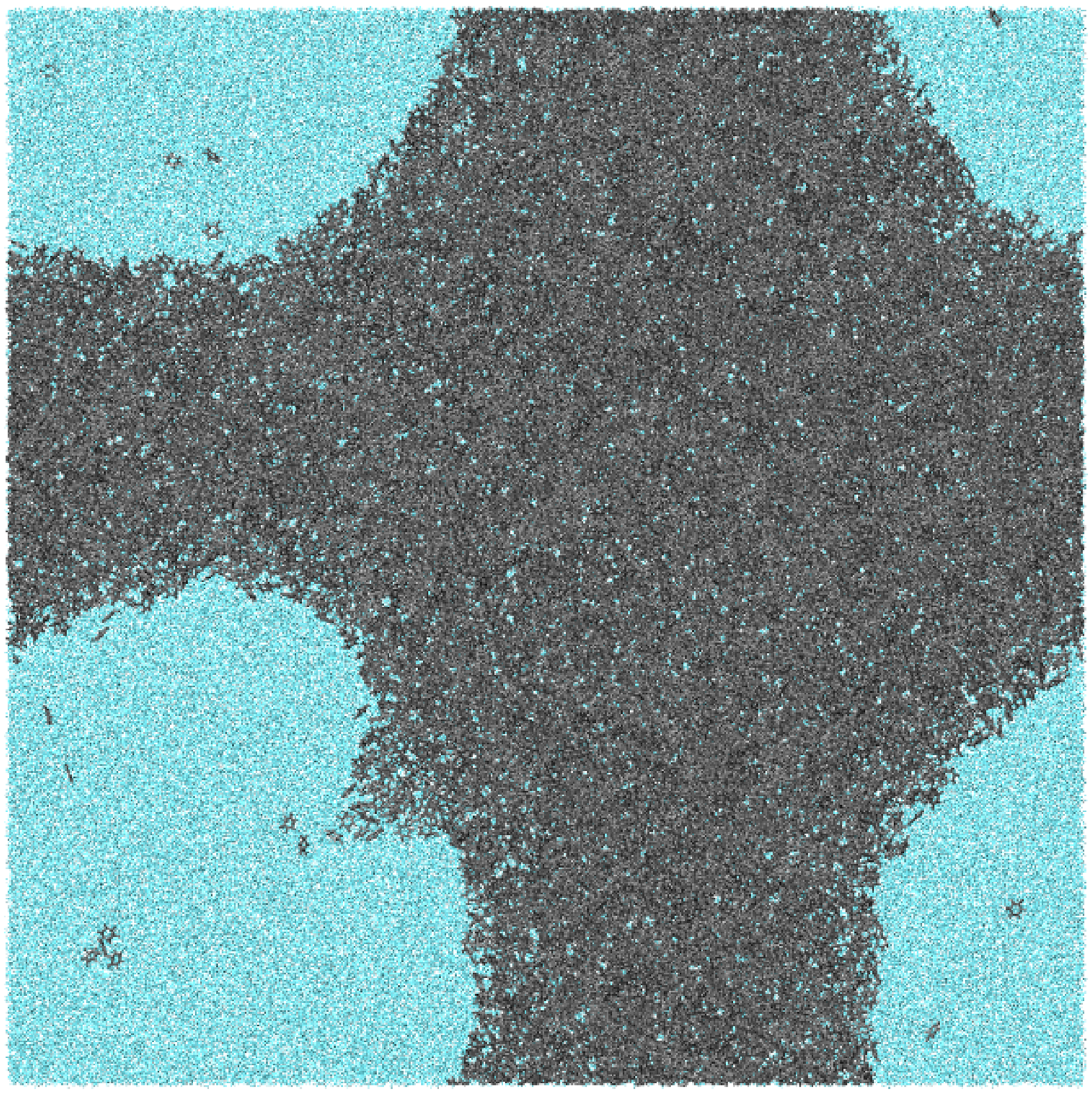}
  \caption{$700\, ps$}
  \label{fig:top_view_700ps}
\end{subfigure}
\begin{subfigure}{.45\textwidth}
  \centering
  \includegraphics[width=.8\textwidth]{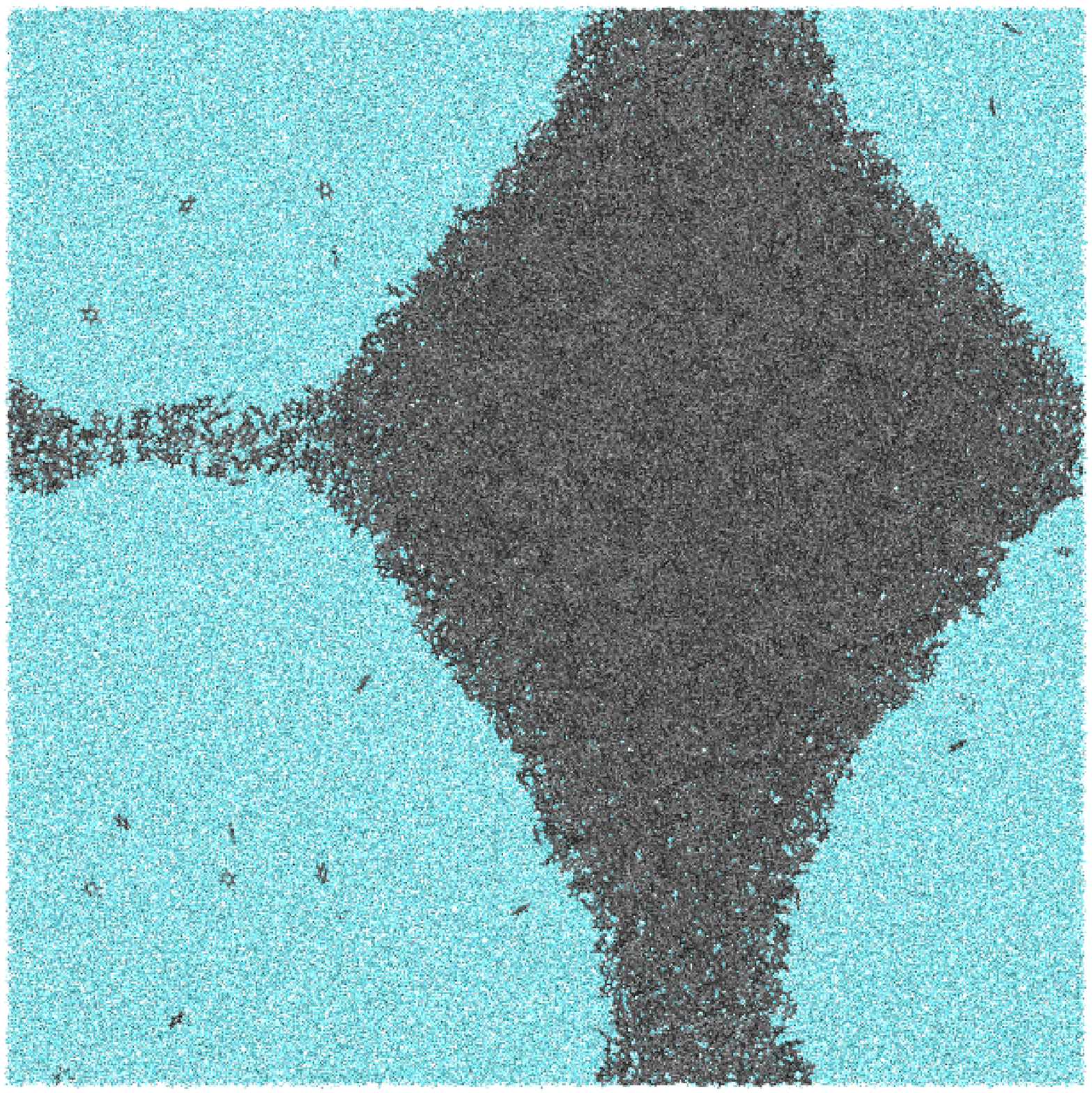}
  \caption{$1500\, ps$}
  \label{fig:top_view_1500ps}
\end{subfigure}
\begin{subfigure}{.45\textwidth}
  \centering
  \includegraphics[width=.8\textwidth]{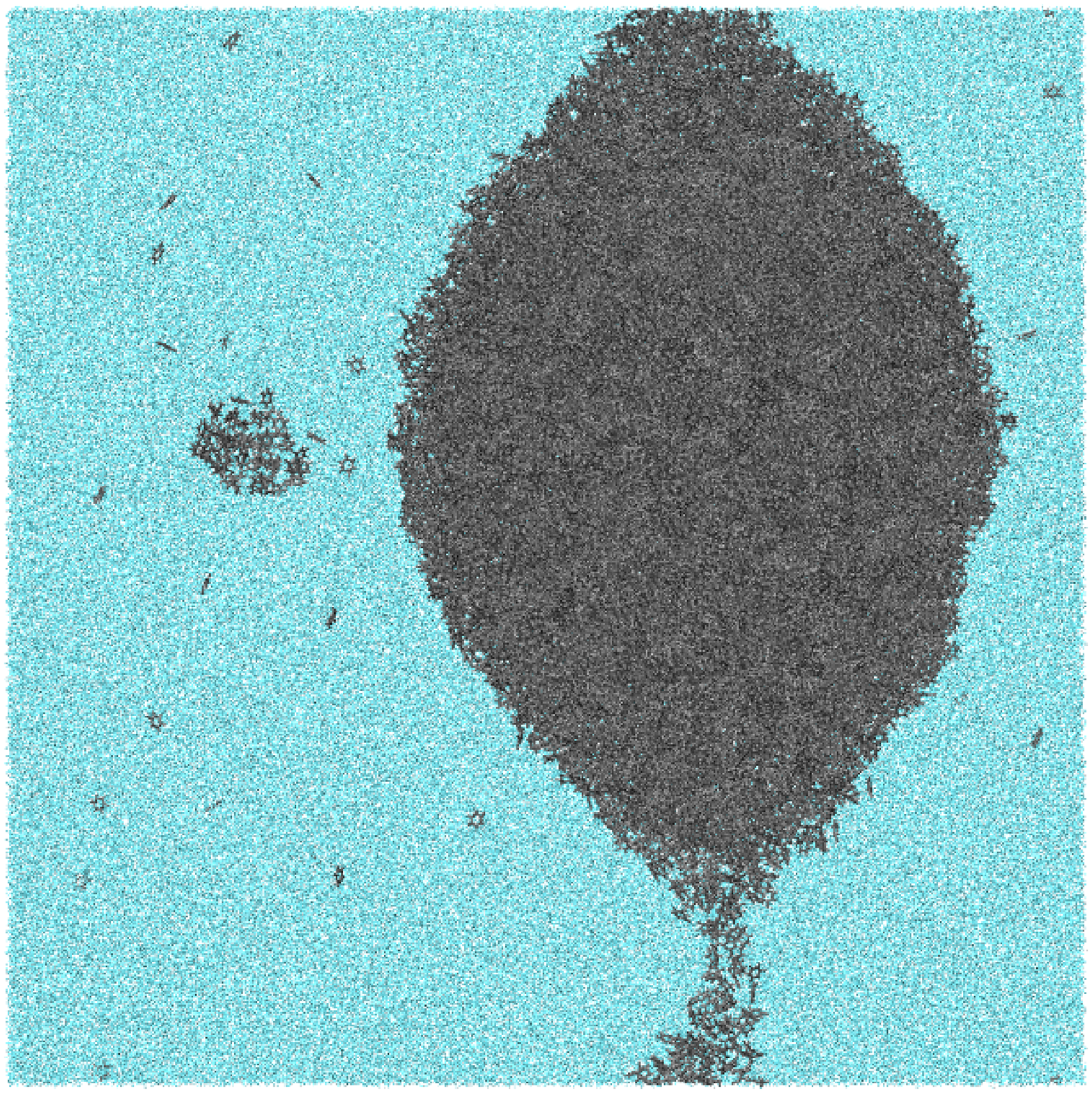}
  \caption{$2000\, ps$}
  \label{fig:top_view_2000ps}
\end{subfigure}
\caption{Top view snapshot of time evolution of film area at applied external electric field of $120\, mV/nm$}
\label{fig:film_rupture_nvt}
\end{figure}

\subsection{$NVT$ simulation - film rupture mechanism}
When the electric field is set to $120\, mV/nm$ a pore is formed in  the film after $500\, ps$. It is observed that the pore expands along the simulation box over time. The time evolution of the pore formation is shown in Figure \ref{fig:film_rupture_nvt} for the time between  $500\, ps$ and $2000\ ps$. The pore is seen as a light spot in Figure \ref{fig:top_view_500ps} at $500\, ps$. When the pore is wide enough water molecules fill in the pore area, seen as a bluish background in the Figure \ref{fig:top_view_700ps} at $700\, ps$. By observing the pore evolution in the toluene film, the time of the complete film rupture (formation of toluene drop)  can be determined.
  

\begin{figure}[h]
\centering
\begin{subfigure}{.32\textwidth}
  \centering
  \includegraphics[width=\linewidth]{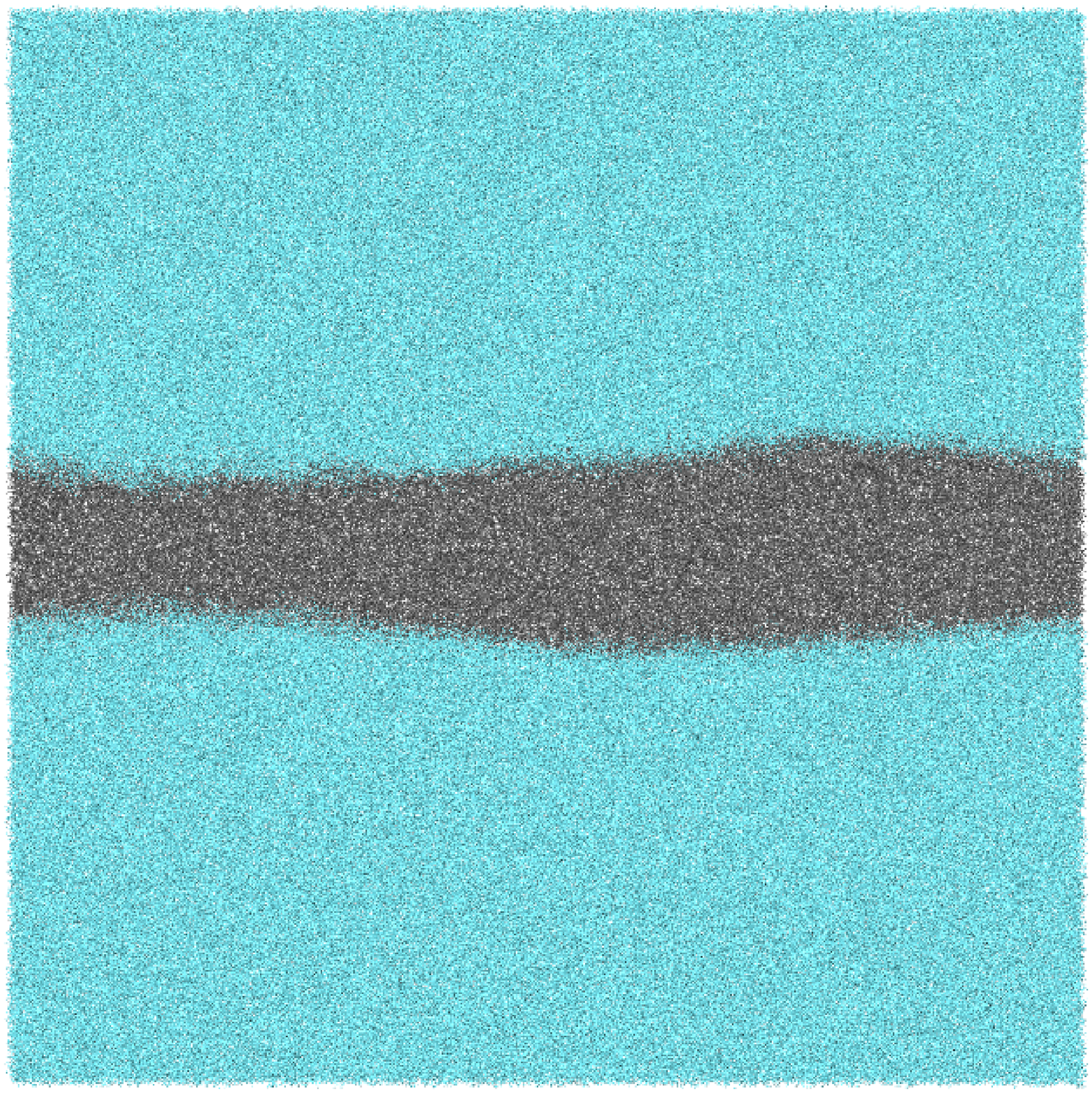}
  \caption{$100\, ps$}
  \label{fig:side_view_100ps}
\end{subfigure}%
\begin{subfigure}{.32\textwidth}
  \centering
  \includegraphics[width=\linewidth]{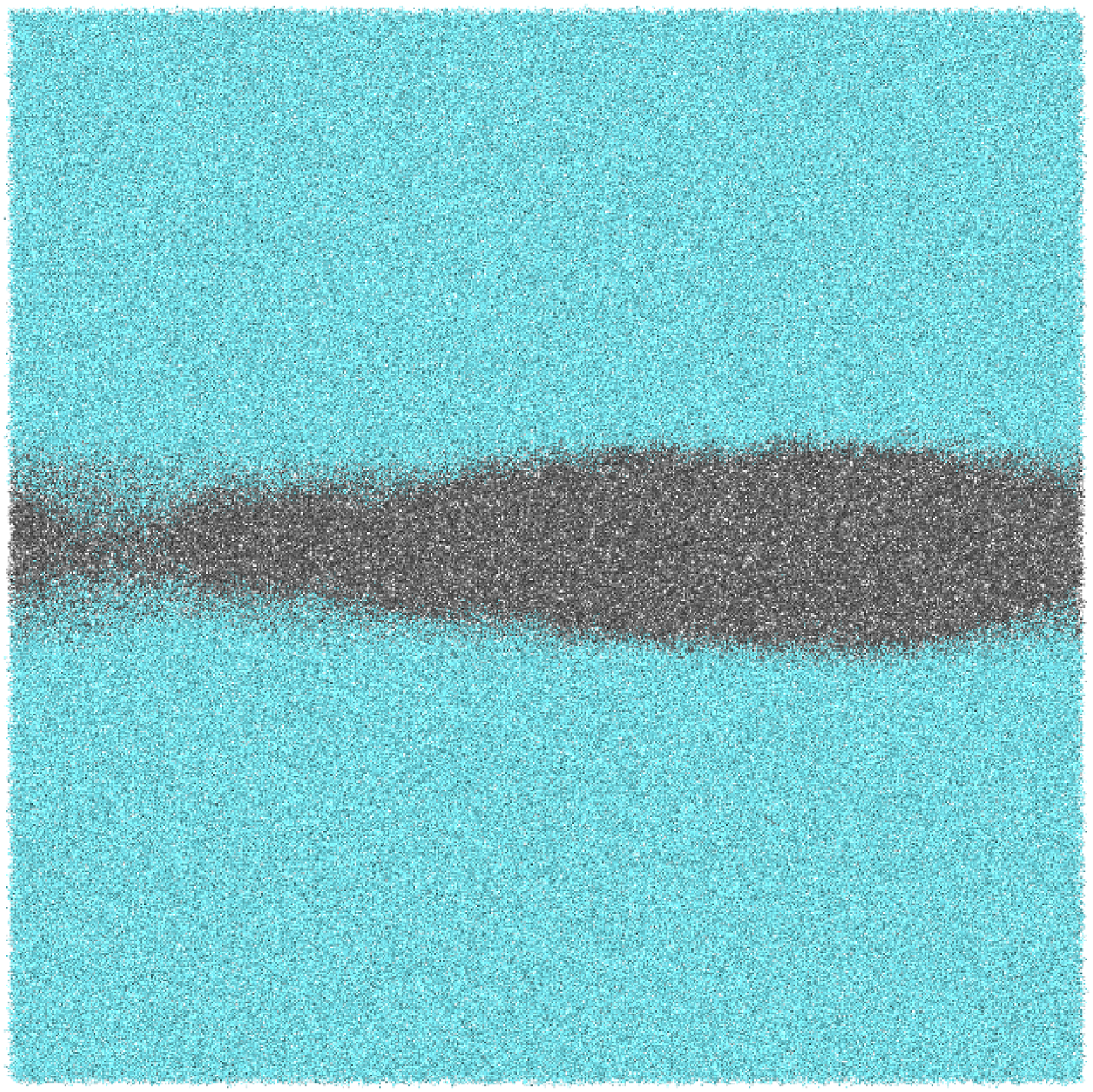}
  \caption{$500\, ps$}
  \label{fig:side_view_500ps}
\end{subfigure}
\begin{subfigure}{.32\textwidth}%
  \centering
  \includegraphics[width=\linewidth]{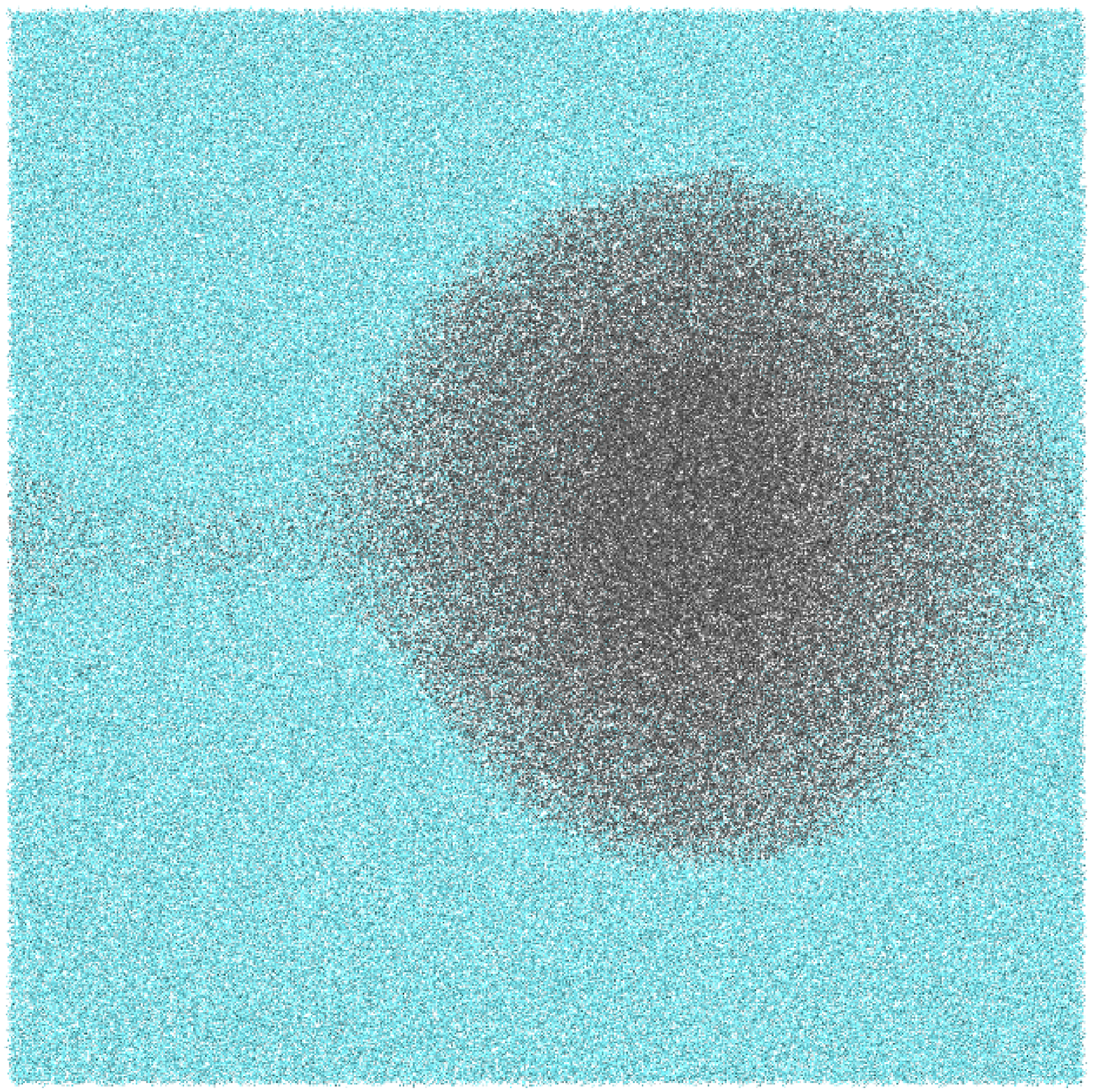}
  \caption{$2000\, ps$}
  \label{fig:side_view_1500ps}
\end{subfigure}
\caption{Side view snapshots for the $120\, mV/nm$ external electric field applied  perpendicularly to the toluene film:  (a) the film profile after $100\, ps$; (b) startup of rupturing of the thinnest part of the film - between $400\, ps$ and $500\, ps$;  (c) breakdown of the film and formation of a toluene drop at about $2000\, ps$}.
\label{fig:nvt_side_view_film_rupture}
\end{figure}

Inspection of the thickness profile at $120 mV/nm$ reveals existence of a dimple inside the toluene film (Figure \ref{fig:side_view_100ps}) prior to the film rupture.The role of the non-homogeneity has been widely reported in thin liquid film literature since 1941 \cite{Landau_1941}. 

Our simulations demonstrate that a non-homogeneous film ruptures at its thinnest place because the electric field strength is the highest there. In other words, the biconcave region of the film interface is subjected to a higher\textendash than\textendash average electrostatic pressure and therefore is a preferred site for the film rupture and nano\textendash pore formation.\\

\subsection{$NVT$ simulation - the film structure}
\begin{figure}[h]
\centering
\begin{subfigure}{.45\textwidth}
  \centering
  \includegraphics[width=\textwidth]{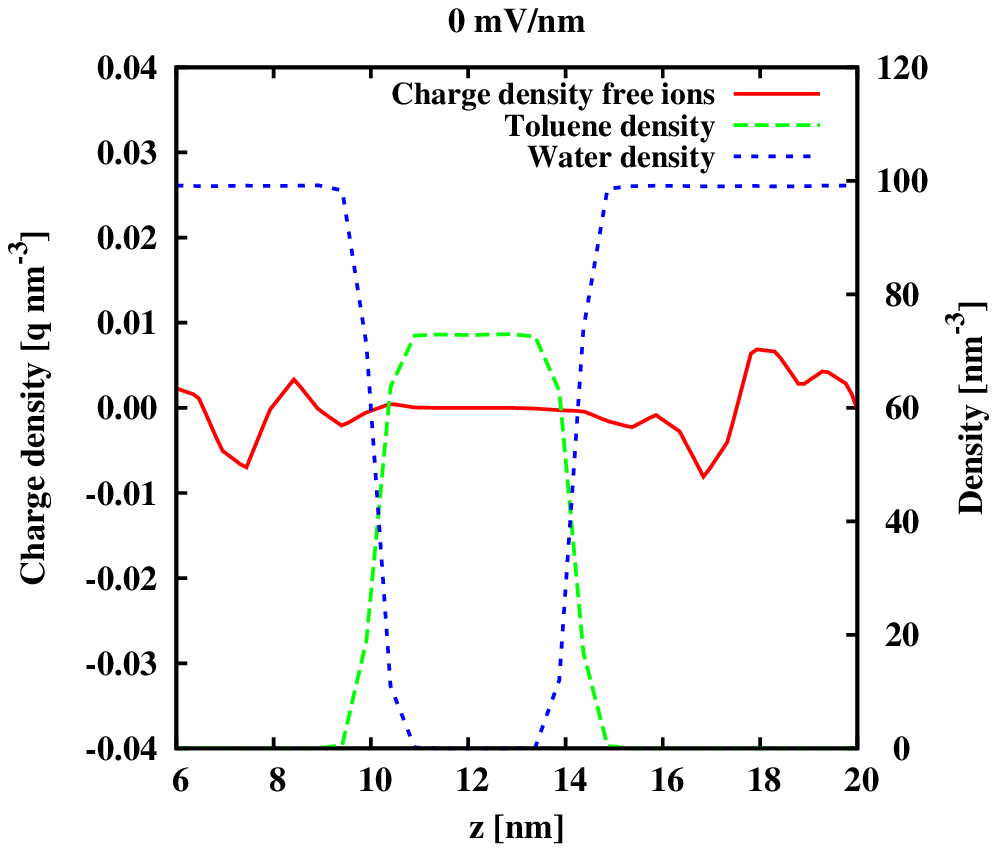}
  \caption{Density distribution at external electric field $0 mV/nm$}
  \label{fig:nvt_dens_dist_0mV}
\end{subfigure}%
\begin{subfigure}{.45\textwidth}
  \centering
  \includegraphics[width=\textwidth]{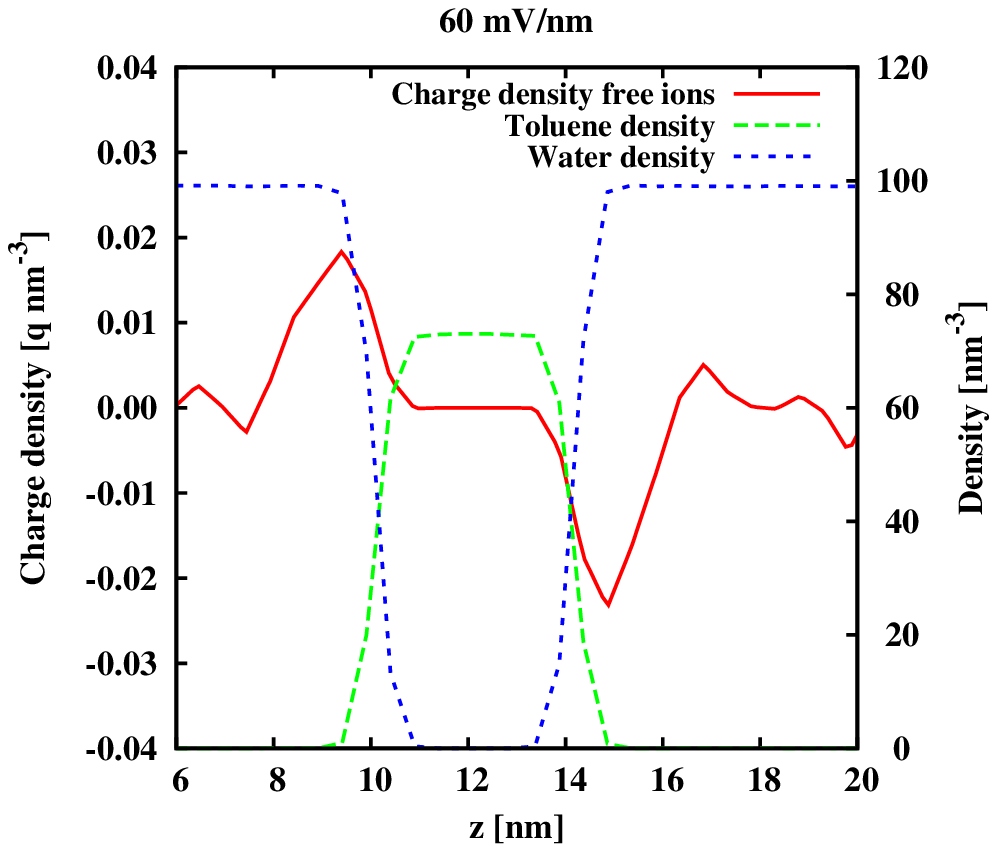}
  \caption{Density distribution at external electric field $60 mV/nm$}
  \label{fig:nvt_dens_dist_60mV}
\end{subfigure}
\begin{subfigure}{.45\textwidth}
  \centering
  \includegraphics[width=\textwidth]{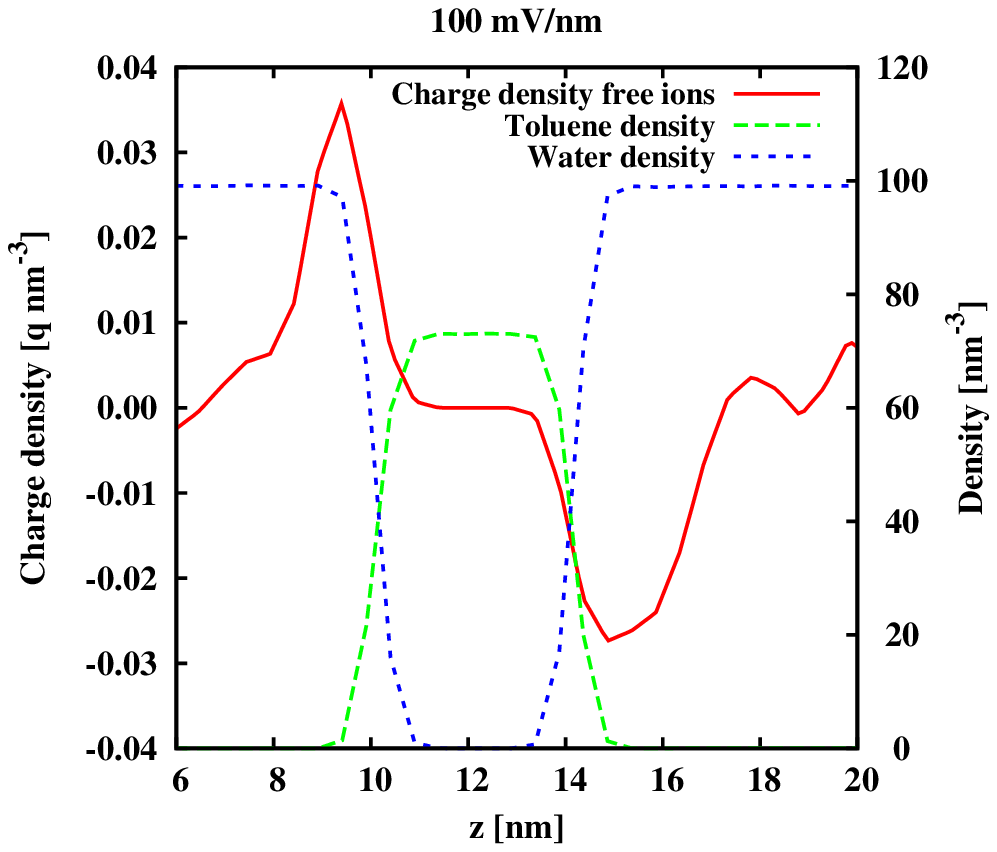}
  \caption{Density distribution at external electric field $100 mV/nm$}
  \label{fig:nvt_dens_dist_100mV}
\end{subfigure}
\begin{subfigure}{.45\textwidth}
  \centering
  \includegraphics[width=\textwidth]{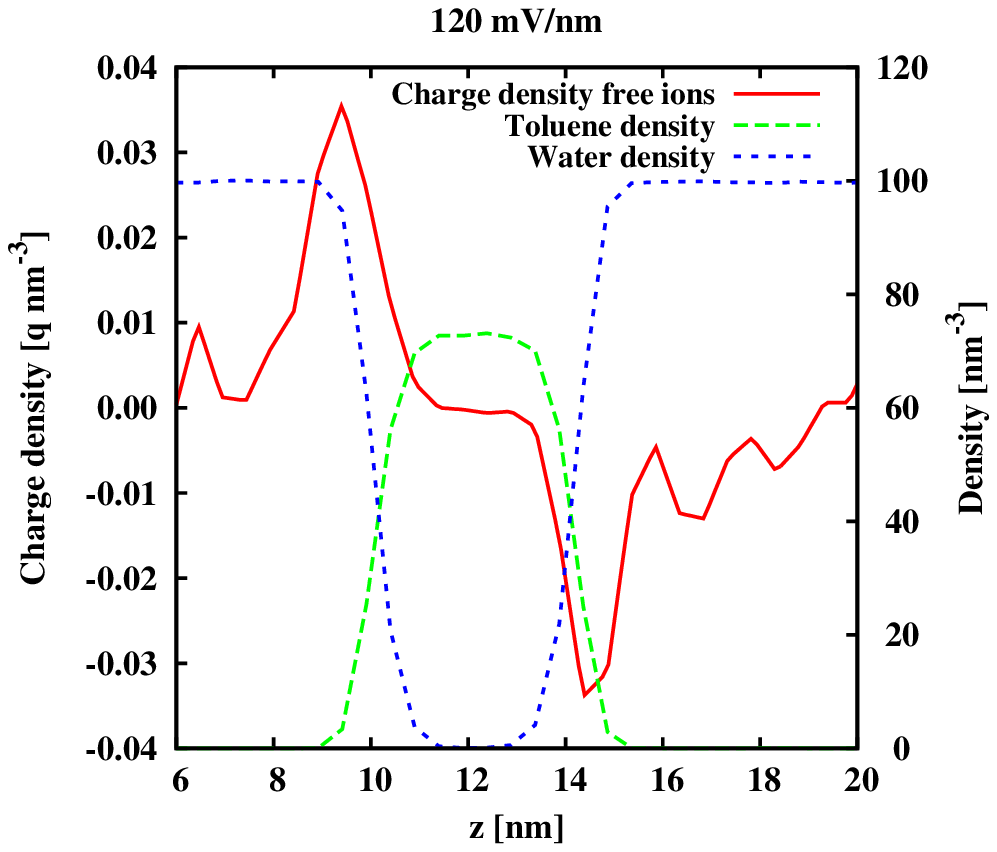}
  \caption{Density distribution at external electric field $120 mV/nm$}
  \label{fig:nvt_dens_dist_120mV}
\end{subfigure}
\caption{Density distribution of free ions (red), toluene (green) and water molecules (blue) at different external electric fields in NVT ensemble}
\label{fig:nvt_all_distributions}
\end{figure}

The charge density  of free ions, water, and toluene molecules in the $z$-direction can be observed in the Figure \ref{fig:nvt_all_distributions}. The  figure illustrates the structure of the toluene film surrounded by aqueous electrolyte solution at $0$, $60$, $100$ and $120$ $mV/nm$ electric field strengths. For the latter case the density was averaged over the first $0.5\, ns$ of simulation, i.e. before the startup of the rupturing process. The part of the  film that contains only toluene molecules is called toluene {\it core}, {\it bulk} water phase is that part of the film, where the density of water molecules is maximal.

The core thickness was estimated using a double\textendash sigmoid function (Equation \ref{eq:double_sigmoid}) at $99\%$ of the plateau height $h$, where $x_0$, $x_1$ being the left and the right half-height respectively, $a$ is the steepness of the sigmoid:
\begin{equation}
f(x;x_0, x_1, a, h) = h\bigg(1 - \frac{1}{1 + \exp^{-a(x-x_1)}}\bigg) \frac{1}{1 + \exp^{-a(x-x_0)}}
\label{eq:double_sigmoid}
\end{equation}

At no applied field, Figure \ref{fig:nvt_dens_dist_0mV} there exists a pure toluene core of $2.6\, nm$ thickness, which is surrounded by two interfacial layers formed by mixing toluene and water interfacial layers. In the toluene interfacial layer, the concentration of toluene molecules decreases in the direction from the pure toluene core towards the bulk water phases, eventually reaching zero concentration. Respectively, in the aqueous interfacial layer, the concentration of water molecules decreases towards the toluene core. Toluene\textendash water mixed layer is $1.4\, nm$ thick, which is estimated as the distance where the toluene density in $z$-direction drops from $99\, \%$ to $1\, \%$ from the plateau height $h$. The bulk aqueous phases contain $Na^+$ and $Cl^-$ ions. The density profiles show that ions penetrate the mixed interfacial layers that border the toluene core. 
Thicknesses of the film core and the interfacial layer are shown in the Table \ref{table:nvt_film_thickness}.

\begin{center}
\begin{table}[ht]
\centering
\begin{tabular}{|c|c|c|c|c|}
\hline
applied field           &$0\, mV/nm$   &$60\, mV/nm$   &$100\, mV/nm$   &$120\, mV/nm$ \\ \hline
toluene core [nm]       &$2.6$         &$2.4$          &$2.0$           &$1.8$   \\ \hline
interfacial layer [nm]  &$1.4$         &$1.7$          &$2.1$           &$2.5$    \\ \hline
total toluene layer [nm]  &$5.4$         &$5.8$          &$6.2$           &$6.8$    \\ \hline
\end{tabular}
\caption{Film thickness at different strength of the applied electric field in NVT ensemble }
\label{table:nvt_film_thickness}
\end{table}
\end{center}

The Figure \ref{fig:nvt_dens_dist_60mV} shows that application of $60\, mV/nm$ field leads to build-up of accumulated positive charges on one side of the toluene film and negative charges on the other side. The peaks of accumulated charges are  situated at the boundary between bulk water and interfacial aqueous layer. In the direction towards toluene core, the concentration of ions decreases  and reaches zero at the toluene core. Ions penetrate only the mixed interfacial region, which is due to the formation of hydration shells. The Figure \ref{fig:nvt_dens_dist_100mV} reveals that field increase up to $100\, mV/nm$ is followed by accumulation of more charges, bringing enough attractive force that leads to thinning of the pure toluene core by $0.6\, nm$ down to $2.0\, nm$, while the thickness of the interfacial layers of toluene and water increases by $0.7\, nm$. 
A possible hypothesis is that the increased electrical compression reshapes the film topography making it a high amplitude rugged surface and the interfacial layer becomes thicker.
The Figure \ref{fig:nvt_dens_dist_120mV} depicts the film structure (profile) at $120\, mV/nm$ and data are averaged over the time interval of $500\, ps$. This  is the moment ($500\, ps$) just before the rupturing process takes place.

The increase of potential from $100$ to $120\, mV/nm$ leads to additional thinning of the toluene core down to $1.8\, nm$. The thickness of interfacial layers of toluene and water increases by $0.4\, nm$. 

\subsection{$NPT$ simulation - the film structure}

\begin{figure}[ht]
\begin{center}
\includegraphics[width=.8\textwidth]{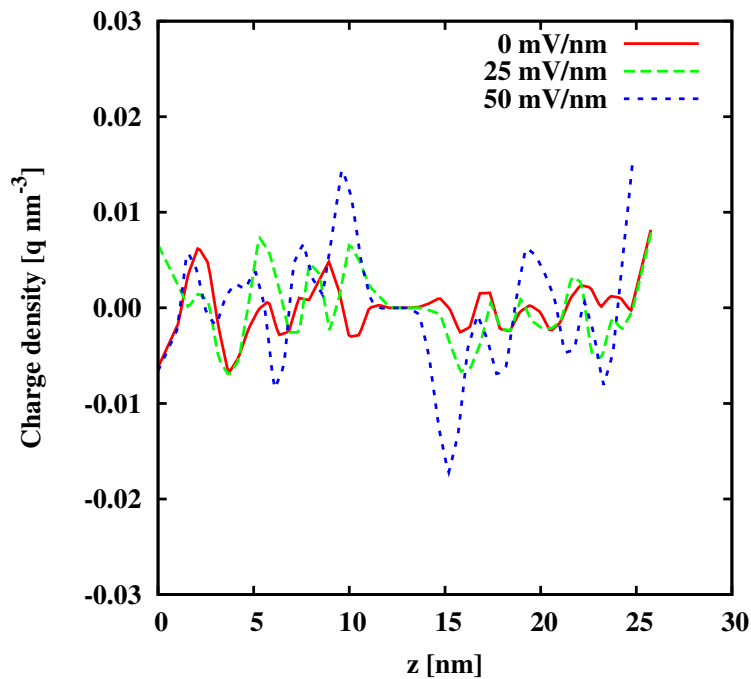}
\end{center}
\caption{Calculated charge of free ions density in $z$-direction for three different external applied electric fields $0$, $25$ and $50$ $mV/nm$ in case of $NPT$ ensemble with constant surface tension.}
\label{fig:npt_charges_all_voltage}
\end{figure}

\begin{figure}[h]
\centering
\begin{subfigure}{.32\textwidth}
  \centering
  \includegraphics[width=\linewidth]{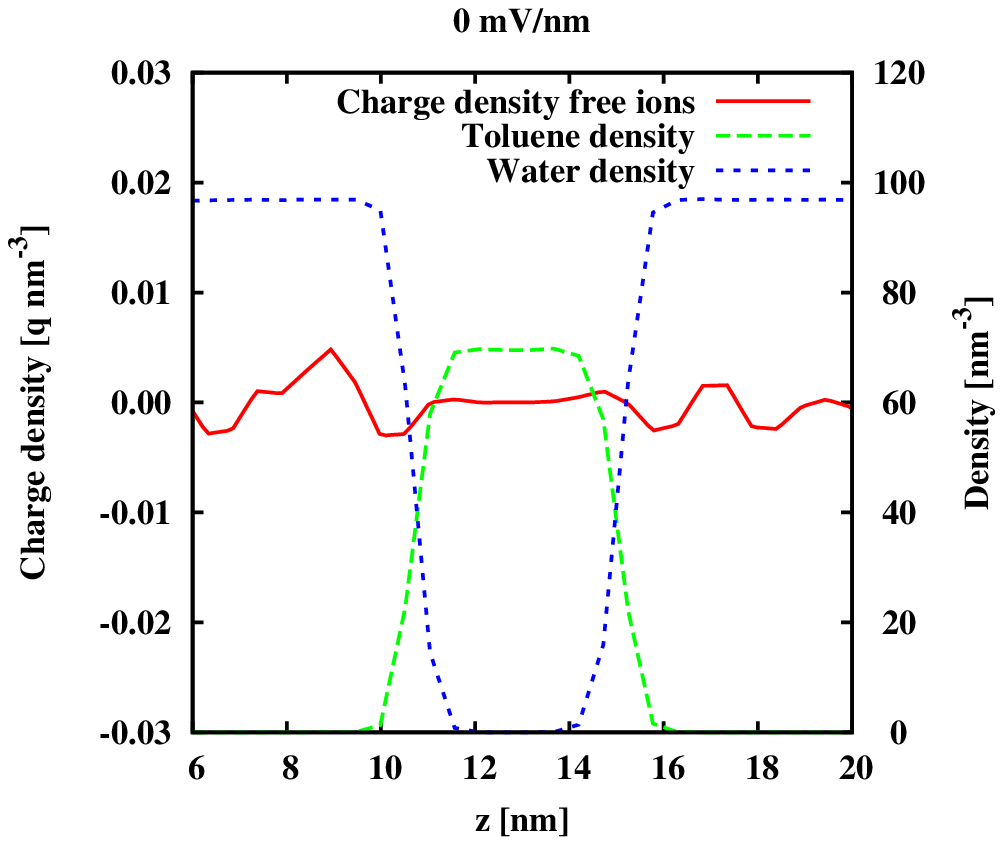}
  \caption{Density distribution at external electric field $0 mV/nm$}
  \label{fig:npt_dens_dist_0mV}
\end{subfigure}%
\begin{subfigure}{.32\textwidth}
  \centering
  \includegraphics[width=\linewidth]{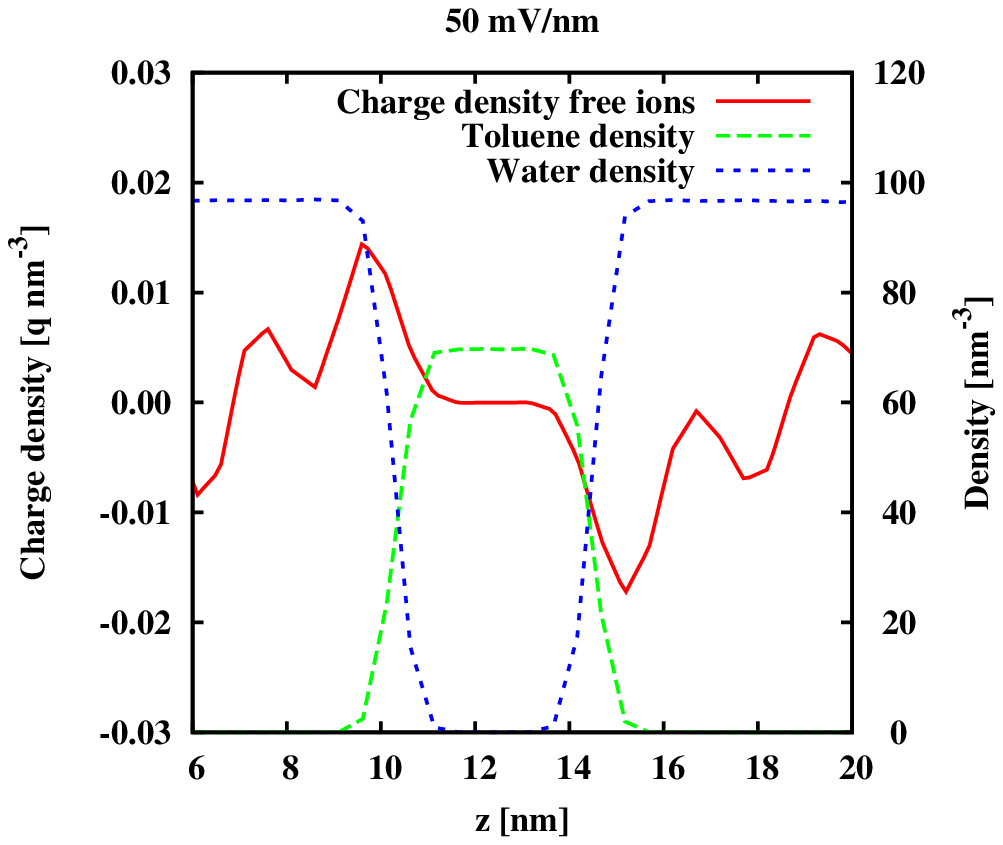}
  \caption{Density distribution at external electric field $50 mV/nm$}
  \label{fig:npt_dens_dist_50mV}
\end{subfigure}
\begin{subfigure}{.32\textwidth}%
  \centering
  \includegraphics[width=\linewidth]{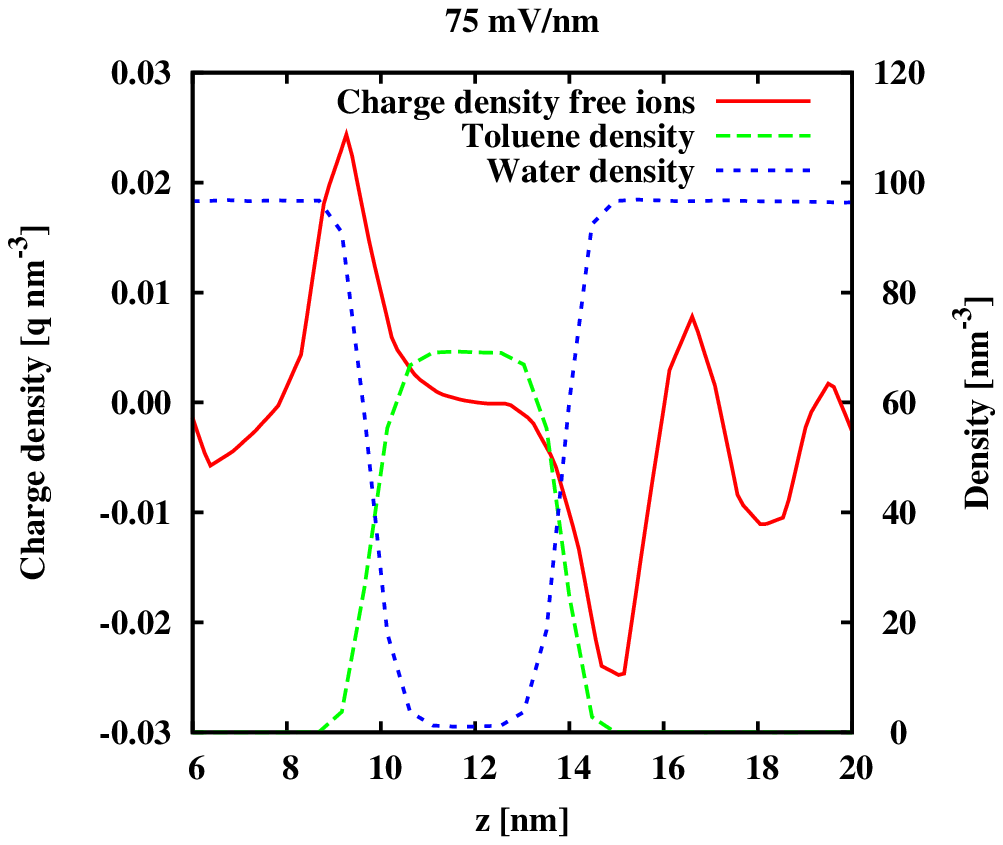}
  \caption{Density distribution at external electric field $75 mV/nm$}
  \label{fig:npt_dens_dist_75mV}
\end{subfigure}
\caption{Density distribution of free ions (red), toluene (green) and water molecules (blue) at different external electric fields in NPT ensemble}.
\label{fig:npt_all_distributions}
\end{figure}

In this section, the results from $NPT$ canonical ensemble simulation are presented. 
The charge  build\textendash up is plotted in the Figure \ref{fig:npt_charges_all_voltage}.
The Figure \ref{fig:npt_all_distributions} depicts the charge build-up at $0$, $25\, mV/nm$,  and $50\, mV/nm$. As in the NVT case, at zero external field no peak formation is observed on both sides of ionic line that exhibits zero-charge density. At $25\, mV/nm$ such formation already takes place and becomes much pronounced at $50$ and $75\, mV/nm$ (Figure \ref{fig:npt_dens_dist_75mV}).  Film rupture occurs at a much lower electric field strength ($75\,  mV/nm$) compared to the NVT simulation ($120\, mV/nm$). Film rupture occurs at a much lower electric field strength ($75\, mV/nm$) compared to the NVT simulation ($120\, mV/nm$). The information regarding the thickness of the toluene core, boundary layers and the total film are summarized in Table \ref{table:npt_film_thickness}. The thickness of different layers were again determined by using the double\textendash sigmoid formula in Equation \ref{eq:double_sigmoid}. At no applied field, the toluene core has the same thickness as in the NVT case. However, the difference between the two simulations is revealed in the size of the mixed boundary zone, being larger for the NPT ensemble. Increase of the electric field, as in the NVT case, again leads to the thinning of the toluene core and to the expansion of the boundary layers. However, at $50\, mV/nm$ (NPT) there is a bit higher thinning of the core, compared to $60\, mV/nm$ (NVT), despite of the lower applied field. This observation, together with the obtained lower critical field could suggest enhanced development of instability in the NPT simulation. It should be noted that in both simulations at the  critical field  the core has the same thickness instants before the film rupture.  Moreover, at that critical field the thickness of the boundary layers and of the total layer are almost identical in both $NVT$ and $NPT$ ensembles.
\begin{center}
\begin{table}[ht]
\centering
\begin{tabular}{|c|c|c|c|}
\hline
applied field           &$0\, mV/nm$   &$50\, mV/nm$   &$75\, mV/nm$   \\ \hline
toluene core [nm]       &$2.6$         &$2.2$          &$1.8$          \\ \hline
interfacial layer [nm]  &$1.8$         &$2.0$          &$2.4$    \\ \hline
total toluene layer [nm]  &$6.2$         &$6.2$          &$6.6$    \\ \hline
\end{tabular}
\caption{Film thickness at different strength of the applied electric field in NPT ensemble }
\label{table:npt_film_thickness}
\end{table}
\end{center}

\section{Conclusions}
The results of MD simulations and their analysis offer insight into the intimate structure of the film, namely the presence of a toluene core, neighboring a mixed boundary zone that contains altogether toluene and water molecules.  Application of external DC electric field leads to redistribution of electrical charges and to the accumulation of oppositely charged ions ($Na^+$ and $Cl^-$) on both sides of the film. Thus, the behavior of the system resembles a liquid capacitor, which charge increases with the rise of the external potential. 
In both $NVT$ and $NPT$ ensembles, {\it condenser plates}, where the charge density is maximal, are situated at the very  border between the bulk aqueous (water) phase and the mixed layer. No ion penetration is observed within the toluene core, thus leaving all the distribution of charges within the mixed zone and the bulk phase that could be attributed to the formation of hydration shells. When critical electric field is reached, within a certain time after the field application electric discharge occurs, indicating the beginning of the rupturing process. Visual snapshots of the evolution of the film area confirm the formation of a hole within the thinnest part of the initially non\textendash homogeneously thin film. 

Results clearly show that in $NPT$ simulations the critical instability is developed at much lower fields (75 $mV/nm$) than in NVT simulations (120 $mV/nm$). First experimental investigation on electro-induced rupture of real toluene-diluted bitumen emulsion films \cite{Panchev200874} shows that critical fields range between 4 and 11 $mV/nm$, depending on the bitumen concentration. Thus, $NPT$ simulation with a constant surface tension appears to be a better choice for further modeling of the systems that resemble more close the real films.
In the $NPT$ ensemble  we can expect that even lower values of the external electric field could rupture the toluene film if we prolong the simulation time. The compressive  action of the built\textendash up charges on both sides of the film is illustrated in the decrease of the thickness of the toluene core with the electric field. The behavior of the system resembles a capacitor with increasing charge with an increase of the external potential. However, in both type of simulations ($NVT$ and $NPT$), the width of the mixed zone and hence of the total film increases with the field increase. 
Moreover, the clarification of the detailed mechanism of the hole formation (wave\textendash like or pore\textendash like) and the role of thickness fluctuations on the rupturing process could be progressed through undertaking an extensive ``subbox'' thickness investigation, when the entire simulation box is divided into boxes along the xy\textendash plane, as each one of them being analyzed. 

In conclusion, we may argue that the model, we have developed for thin films, provides a ground for implementing a further complication of the investigated system, introducing surface active molecules, as well as a verification of our expectances for a decrease of the critical electric field when longer simulations are performed.






\section{Acknowledgements}
S.Pisov and D. Dimova acknowledge the access to the HPC cluster in Sofia Tech Park, used for the heavy computations. S. Madurga acknowledges financial support from the Generalitat de Catalunya (grant 2014-SGR-1251). The support of H2020 program of the European Union (project Materials Networking) is gratefully acknowledged by S. Madurga and M. Nedyalkova.


%

%
%

\bibliographystyle{natbib}
\bibliography{references}

\end{document}